# 3-phonon scattering pathways for vibrational energy transfer in crystalline RDX

Gaurav Kumar[1], Francis G. VanGessel[2], Lynn B. Munday[3], Peter W. Chung*[4]

KEYWORDS. energetic materials, RDX, phonon scattering, vibrational energy transfer

## Abstract

A long-held belief is that shock energy induces initiation of an energetic material through an indirect energy up-pumping mechanism involving phonon scattering through doorway modes. In this paper, a 3-phonon theoretical analysis of energy up-pumping in RDX is presented that involves both direct and indirect pathways where the direct energy transfer dominates. The calculation considers individual phonon modes which are then analyzed in bands. Scattering is handled up to the third order term in the Hamiltonian based on Fermi's Golden Rule. On average, modes with frequencies up to 90 cm$^{-1}$ scatter quickly and redistribute the energy to all the modes. This direct stimulation occurs rapidly, within 0.16 ps, and involves distortions to NN bonds. Modes from 90 to 1839 cm$^{-1}$ further up-pump the energy to NN bond distortion modes through an indirect route within 5.6 ps. The highest frequency modes have the lowest contribution to energy transfer due to their lower participation in phonon-phonon scattering. The modes stimulated directly by the shock with frequencies up to 90 cm$^{-1}$ are estimated to account for 52 to 89% of the total energy transfer to various NN bond distorting modes.

[1] Department of Mechanical Engineering, University of Maryland in College Park, Maryland, USA

[2] US Naval Surface Warfare Center Indian Head Division, Indian Head, Maryland, USA

[3] US Department of Energy, Idaho National Laboratory, Idaho Falls, Idaho, USA

[4] Department of Mechanical Engineering, University of Maryland in College Park, Maryland, USA, pchung15@umd.edu

## 1. Introduction

Energetic materials (EMs) appear in a wide range of industrial and defense applications. Understanding their physical and chemical processes is critical for their development, safe usage, transport and storage. Considerable efforts have studied initiation mechanism in EMs where mechanical energy input creates conditions in the material that make it more susceptible to bond rupture and possible self-sustaining chemical reactions. Bowden and Yoffe [1, 2, 3] showed that the bulk heating of the material in shocked energetic crystals is insufficient to describe the initiation processes. As a result, the concept of localization of energy into highly concentrated regions, called hotspots, was developed [2, 4, 5, 6]. Subsequently, to explain the localization of energy, an indirect mechanism of energy transfer via multi-phonon up-pumping was postulated [7, 8, 9, 10]. This model assumed that mechanical energy from the shock excites the low frequency phonons which results in a rapid increase in their population. The energy from these low frequency modes is transferred into the mid-frequency modes, referred to as doorway modes, via phonon-phonon scattering. The energy from the doorway modes is, in turn, transferred into the high frequency intramolecular vibrations which is believed to increase the population and amplitude of internal molecular vibrations beyond bond dissociation limits. The indirect energy transfer mechanism meant that the mid frequency doorway modes are critical for phenomena leading to initiation in energetics. The model was further extended by Dlott and co-workers who studied terahertz excitation in molecular solids, measured by ps timescale vibrational spectroscopy, and included a description of localization of energy at crystal defects [11, 12, 13, 14]. Dlott et al. concluded that the rate of up-pumping of energy depends on the number of doorway modes and the Gruneisen parameter values which are indicative of the strength of anharmonic coupling between the modes [15].

For the indirect up-pumping model for initiation to work, strong scattering must occur among a) modes at the low frequencies close to the frequencies associated with mechanical shock, b) the frequencies of the modes believed to serve as the doorway modes, and c) the modes associated with the largest distortions of the bonds, so-called vibrons, likely to possess dissociation energies which, when released, can sustain exothermic reactions. Aubuchon et al. [16], in their work on modes that result in asymmetric stretching of the nitro functional group ($NO_2$) in TNAX, RDX, HMX, and CL-20, suggested that the relaxation of the intra-molecular vibrons does not involve the low frequency phonon modes. Other works have also posited that the energy transfer process may occur through a direct route simply due to the NN activity found in low frequency eigenmodes without intermediate energy transfer or the involvement of doorway modes [17]. A definitive understanding of phonon mechanisms in energetic materials therefore still remains elusive. It requires a detailed determination of vibrational energy transfer not only of the relaxation behavior of individual modes, but a complete picture of mode-to-mode scattering that may drive energy through the intrinsic scattering network.

Much progress has occurred toward this goal motivated by early interest in the relationship between vibrational energy transfer and sensitivity. To calculate the total energy transfer rate into vibron bands, Fried and Ruggiero [18] derived a simple formula in terms of the density of vibrational states and the vibron-phonon coupling, which were calculated using existing inelastic neutron scattering data. They studied TATB, $\gamma$ and $\beta$-HMX, RDX, Pb-styphnate, Styphnic acid and Picric acid, and observed that the estimated energy transfer rates in pure unreacted material are several times greater in sensitive explosives than in insensitive explosives. Following Fried and Ruggiero's formula, Koshi and co-workers investigated a broad range of EMs including PETN, HMX, RDX, Tetryl, TNT, FOX-7, ANTA, PN, NQ and DMN, and observed a good correlation of the energy transfer rates with the impact sensitivity of the explosives [19, 20, 21]. Similar observations were made by Bernstein [22], Joshi et al. [23], and McNesby et al. [24] who investigated the rate of energy transfer from the low frequency phonons to the higher frequency vibrons. Aubuchon et al. used IR pump-probe spectroscopy to show rapid relaxation (2 to 6 ps time scale) of the asymmetric stretching mode of the nitro functional group ($NO_2$) of several molecules used in TNAZ, RDX, HMX and CL-20 [16]. Ostrander et al. also used IR pump-probe spectroscopy to study the asymmetric stretch of the nitrate ester groups in PETN thin films and suggested an energy transfer pathway with a 2 ps time scale [25]. Numerous other works have explored the role of various crystal distorting modes in the transfer of energy in EMs [17, 26, 27, 28, 29, 30, 31, 32]. These works provide useful insights into the phonon modes which can be critical in phenomena leading to initiation and have motivated investigation of the critical bond stretching and bending modes.

A significant challenge in the development of a complete phonon picture of up-pumping has been the lack of a quasiparticle model that accurately accounts for third (and higher) order terms in the Hamiltonian vis-à-vis Fermi's Golden Rule [33] across a larger range of the Brillouin zone. But with the perpetual development of advanced computational architectures, efforts have been made to calculate the anharmonic phonon properties more accurately to model the energy transfer between the vibrational modes. Early efforts were based on approximations to the anharmonic terms that made it possible to overcome the computational costs at the expense of making it difficult to generalize the consideration of distinct phonon modes. Hooper developed a parametric expression for transition probabilities for nitromethane, RDX, and HMX that require assumptions including SMRTA, the applicability of Debye's model for the pDOS, the occupation numbers being independent of frequency, and that only modes of equal frequency can scatter [34]. Long and Chen calculated phonon-phonon scattering rates in TATB by developing a stress-frequency relationship as a means of representing anharmonic effects without the high computational cost of a 3-phonon scattering Hamiltonian [35]. Michalchuk et al. studied the sensitivity of a wide range of EMs using a 3-phonon scattering model derived from Fermi's Golden Rule (FGR) [36]. This work was built on the so-called average anharmonic approximation [37], which reduces computational cost by assuming that all elements of the cubic anharmonic matrix $V^{(3)}$ are equal. Recently, however, estimates of thermal properties based on quasiparticle models have shown that a more complete representation of carriers in the Brillouin zone may be necessary for accurate modeling of phonon mechanisms. Kumar and coworkers showed that thermal conductivity of energetic crystals like RDX can be dominated by non-acoustic carriers, which constitute the majority of the phonon modes [38], and that modes outside of the acoustic bands can contribute substantially to NN and other intramolecular bond distortions [39, 40]. These observations suggest that the use of approximations that otherwise reduce the complexity of the rich set of optical modes in these materials may omit important contributions from a large number of modes to the transfer of vibrational energy.

In this work, the perturbation theory is used to calculate the 3-phonon scattering rates and the mode-to-mode scattering rate matrix for all branches in the energetic material Cyclotrimethylenetrinitramine ($C_3H_6N_6O_6$, also commonly known as RDX) under ambient conditions. The role of all phonon modes in the transfer of vibrational energy is investigated which indicates a hybrid mechanism of energy up-pumping to the critical NN bond distortion modes via both direct and indirect pathways where the direct route is faster and accounts for the majority of the energy transfer.

## 2. Computational Details

### 2.1 Energy Minimization, Harmonic Properties and Force Constants

The equilibrium minimum energy structure of RDX in [41] is used as the initial configuration in this work. A flexible molecule potential is used for all calculations [42]. The minimum free energy structure of RDX at 300 K is calculated using the open source package GULP [43] which minimizes the Helmholtz free energy $A = U_{static} + U_{vib} - TS_{vib}$ [44], where $U_{static}$ is the static internal energy that would be calculated in a conventional energy minimization, $U_{vib}$ is the vibrational energy, $T$ is the temperature and $S_{vib}$ is the vibrational entropy. The phonon mode frequencies and eigenvectors are also determined using GULP for a uniform $6 \times 6 \times 6$ k-points. This results in a total of (504 branches) × (216 k-points) = 108864 phonon modes.

The open source packages LAMMPS [45] and ShengBTE [46] are used to calculate the third order interatomic force constants (IFCs), the 3-phonon scattering rates and mode-to-mode scattering rates. LAMMPS determines the single point energies which are then provided to ShengBTE to perform the finite differences needed to calculate the third derivatives of energy for the IFCs. The IFCs are determined over a symmetry-reduced crystal. Finally, using the phonon frequencies, eigenvectors and IFCs, the scattering rates are calculated for all modes.

For ease in discussing the results, the spectrum is divided into ten frequency bands (I through X), as shown in Table 1. The frequency ranges of the bands are defined roughly based on groupings of modes in contiguous branches according to the way in which those modes distort the crystal. This is done by identifying for each mode the atoms in the unitcell with the largest displacement. The calculation details of the displacement of atoms and distortion of bonds for each mode are presented in Section 1 of the Supporting Information. The modes in contiguous branches with the same largest displaced atoms and similar frequency are grouped together. In addition, the modes from 0 to 90 cm$^{-1}$ that correspond to translation of the molecules are divided into band I (0 to 30 cm$^{-1}$), band II (30 to 50 cm$^{-1}$) and band III (50 to 90 cm$^{-1}$) to investigate the relative importance of these three bands in the energy transfer mechanism. The qualitative nature of the distortions associated with the modes in each band are listed in Table 1. For reference, the assignment of vibrational modes of RDX found in other experimental literature is also listed in Table 1. It should be noted that as a result of this approach to parsing the frequency bands, the number of phonon modes in different bands are different, depending on the density of states and the frequency range of individual bands.

| Bands | Frequency range (cm$^{-1}$) | Atom(s) with the largest displacement | Largest bond, angle, or dihedral distortions | Experiment Literature |
|---|---|---|---|---|
| I | 0 to 30 | Translation | - | - |
| II | 30 to 50 | Translation | - | - |
| III | 50 to 90 | Translation | - | - |
| IV | 90 to 139 | NO$_2$ | NN, NO, NNC, ONNC, HCNC | NO$_2$ [47] |
| V | 144 to 262 | O | NO, ONO, NNC, NOON, ONNC | NO$_2$ [47], N-NC [26] |
| VI | 352 to 502 | C and H | CH, CN, NNC, NCN, ONNC, HCNC | Ring bending [47] |
| VII | 541 to 732 | N-NO$_2$ | NN, NO, ONO, ONN, NOON, ONNC | N-NO$_2$ [26, 48] |
| VIII | 930 to 1839 | C and H | CH, CN, NCH, HCH, ONNC, CNCN | C-N, CH$_2$ [26, 47] |
| IX | 2055 to 2489 | NO$_2$ | NN, NO, ONO, ONN, ONNC, HCNC | - |
| X | 3342 to 3444 | C and H | CH, HCH, NCH, ONNC, CNCN | C-H [47, 48, 49] |

**Table 1.** Frequency range of ten phonon bands, corresponding atoms with the largest displacement, and bonds with the largest stretching and bending in the unitcell.

## 2.2 Mode-to-mode Phonon Scattering Rates

The crystal Hamiltonian can be written as

$$H = H_0 + H_3 + H_4 \ldots \quad (1)$$

where $H_0$ is the harmonic term, $H_3$ and $H_4$ are anharmonic terms also referred to as first and second order perturbation terms respectively. In this work, we only consider up to the first order perturbation term of the Hamiltonian. The first order perturbation term $H_3$ is defined as [50, 51]

$$H_3 = H^{(3)}_{\phi_1,\phi_2,\phi_3}(a^\dagger_{-\phi_1} + a_{\phi_1})(a^\dagger_{-\phi_2} + a_{\phi_2})(a^\dagger_{-\phi_3} + a_{\phi_3}) \quad (2)$$

where $\phi$ is the phonon mode index ($\phi_1, \phi_2, \phi_3$ are mode indices of the three phonons involved in scattering, $-\phi$ refers to a mode corresponding to a negative wavevector), $H^{(3)}_{\phi_1,\phi_2,\phi_3}$ are Fourier transforms of the third order IFCs, $a^\dagger_\phi$ and $a_\phi$ are creation and annihilation operators respectively with $a^\dagger_\phi|n_\phi\rangle = \sqrt{n_\phi + 1}\,|n_\phi + 1\rangle$ and $a_\phi|n_\phi\rangle = \sqrt{n_\phi}\,|n_\phi - 1\rangle$, and $n_\phi$ is the phonon mode population. With the above expression for the anharmonic Hamiltonian, Maradudin and Fein formulated a method to calculate the intrinsic phonon scattering rates using the perturbation theory (Fermi's Golden Rule or FGR) [50]. Details of the scattering rate calculations can be found in Section 2 of the Supporting Information. Using FGR under single mode relaxation time approximation (SMRTA), which assumes that the non-equilibrium population of any mode is calculated independently of other phonon modes i.e., $n_{\phi_1} = n^0_{\phi_1} + n'_{\phi_1}$ and $n_{\phi_2} = n^0_{\phi_2}$, $n_{\phi_3} = n^0_{\phi_3}$, the rate of change of occupation of the mode $\phi_1$ can be expressed as

$$\frac{\partial n_{\phi_1}}{\partial t} = -n'_{\phi_1} \sum_{\phi_2,\phi_3} \left\{ \frac{1}{2}(1 + n^0_{\phi_2} + n^0_{\phi_3})L_- + (n^0_{\phi_2} - n^0_{\phi_3})L_+ \right\} \quad (3)$$

where $n'_{\phi_1}$ is the perturbation in population of the mode $\phi_1$, $n^0_\phi$ is the equilibrium phonon population of the mode $\phi$ given by Bose-Einstein (BE) statistics, and the summation on the right side is the intrinsic 3-phonon scattering rate $(\Gamma_{\phi_1})$,

$$\Gamma_{\phi_1} = \sum_{\phi_2,\phi_3} \left\{ \frac{1}{2}(1 + n^0_{\phi_2} + n^0_{\phi_3})L_- + (n^0_{\phi_2} - n^0_{\phi_3})L_+ \right\} \quad (4)$$

where $L_\pm = \frac{\pi\hbar}{4N}\left|V^{(3)}_\pm\right|^2 \Delta_\pm \frac{\delta(\omega_{\phi_1} \pm \omega_{\phi_2} - \omega_{\phi_3})}{\omega_{\phi_1}\omega_{\phi_2}\omega_{\phi_3}}$ account for the conservation of crystal momentum and energy and the probability of transition from an initial state to a final state for absorption (represented by $-$) and emission (represented by $+$) processes. The scattering rate ($\Gamma_{\phi_1}$) can be considered to be a product of two terms, $\left|V^{(3)}_\pm\right|^2$ which indicates the anharmonicity of the modes and a 3-phonon phase space volume $P_{3,\phi_1}$ calculated as

$$P_{3,\phi_1} = \sum_{\phi_2,\phi_3} \left\{ \frac{1}{2}(1 + n^0_{\phi_2} + n^0_{\phi_3})\Delta_- \frac{\delta(\omega_{\phi_1} - \omega_{\phi_2} - \omega_{\phi_3})}{\omega_{\phi_1}\omega_{\phi_2}\omega_{\phi_3}} \right. \\ \left. + (n^0_{\phi_2} - n^0_{\phi_3})\Delta_+ \frac{\delta(\omega_{\phi_1} + \omega_{\phi_2} - \omega_{\phi_3})}{\omega_{\phi_1}\omega_{\phi_2}\omega_{\phi_3}} \right\} \quad (5)$$

which indicates the number of allowed 3-phonon scattering events that follow conservation of crystal momentum and energy. The phonon lifetime ($\tau_{\phi_1}$) is defined as the inverse of the scattering rate,

$$\tau_{\phi_1} = \frac{1}{\Gamma_{\phi_1}} \quad (6)$$

The present approach uses 216 discrete points (64 after symmetry is taken into consideration) in the Brillouin zone for every branch in the spectrum. Due to the present consideration of these discrete phonon modes, each term in the sum in Eq. (4) can be evaluated and examined. Namely, we can breakdown the sum $\Gamma_{\phi_1} = \sum_{\phi_2} \Gamma_{\phi_1,\phi_2}$ to understand the mediating role of all other modes on the relaxation of a given mode $\phi_1$. Thus, the mode-to-mode scattering rate (contribution of the mode $\phi_2$ to relaxation of the mode $\phi_1$) is

$$\Gamma_{\phi_1,\phi_2} = \sum_{\phi_3} \left\{ \frac{1}{2}(1 + n^0_{\phi_2} + n^0_{\phi_3})L_- + (n^0_{\phi_2} - n^0_{\phi_3})L_+ \right\} \quad (7)$$

This definition provides the probability of scattering between any two modes, $\phi_1$ and $\phi_2$, as mediated by *all* allowable scattering events involving mode $\phi_3$ via 3-phonon processes. Since the absorption and emission terms constitute both creation and annihilation events, the probability measures the likelihood of $\phi_1$ and $\phi_2$ scattering but does not indicate directionality. Furthermore, due to the mediating role of $\phi_3$ and the conservation rules, the scattering rates are not necessarily symmetric with respect to $\phi_1$ and $\phi_2$; namely the effect of scattering involving $\phi_2$ on the phonon occupation level of $\phi_1$ is unequal to the scattering involving $\phi_1$ on the occupation level of $\phi_2$. Upon grouping the modes into bands, the average percent contribution of any $band2$ to the scattering rate of any $band1$, which we refer to as band-to-band scattering, is calculated as

$$\Gamma_{band1,band2} = \frac{1}{\text{no. of modes in } band1} \sum_{\phi_1 \in band1} 100 \frac{\sum_{\phi_2 \in band2} \Gamma_{\phi_1,\phi_2}}{\Gamma_{\phi_1}} \qquad (8)$$

and the contribution of individual modes to the scattering of NN bands (bands IV, VII and IX) $\Gamma_{NN\ bands,\phi_2}$ is calculated as

$$\Gamma_{NN\ bands,\phi_2} = \sum_{\phi_1 \in band\ IV, VII, IX} \Gamma_{\phi_1,\phi_2} \qquad (9)$$

Next, we derive the relationship between energy transfer rate and scattering for any given mode $\phi_1$ for small perturbation in phonon population of the modes. The energy transfer rate can be expressed as

$$\frac{\partial E_{\phi_1}}{\partial t} = \hbar\omega_{\phi_1} \frac{\partial n_{\phi_1}}{\partial t} \qquad (10)$$

where the partial derivative on the right side can be obtained by solving the Boltzmann Transport Equation (BTE) which, in the absence of an external force, is expressed as [52]

$$\frac{\partial n_{\phi_1}}{\partial t} + v_{\phi_1} \cdot \nabla n_{\phi_1} = \left(\frac{dn_{\phi_1}}{dt}\right)_{scattering} \qquad (11)$$

where $v_{\phi_1}$ is the phonon mode group velocity and the term on the right side is the rate of change in phonon population due to scattering as calculated in Eq. (3). Under the approximation of small perturbation i.e., $\Delta n_{\phi_1} \to 0$, Eq. (11) reduces to $\frac{\partial n_{\phi_1}}{\partial t} = \left(\frac{dn_{\phi_1}}{dt}\right)_{scattering}$. Using this result and substituting Eq. (3) into Eq. (10), the energy transfer rate for mode $\phi_1$ is calculated as

$$\frac{\partial E_{\phi_1}}{\partial t} = -n'_{\phi_1} \hbar \omega_{\phi_1} \Gamma_{\phi_1} \tag{12}$$

Similarly, the contribution of any mode $\phi_2$ to energy transfer rate of mode $\phi_1$ is calculated as

$$\frac{\partial E_{\phi_1,\phi_2}}{\partial t} = -n'_{\phi_1} \hbar \omega_{\phi_1} \Gamma_{\phi_1,\phi_2} \tag{13}$$

For a given mode $\phi_1$, we see that $\frac{\partial E_{\phi_1,\phi_2}}{\partial t}$ is directly proportional to $\Gamma_{\phi_1,\phi_2}$. Therefore, the degree of contribution of mode $\phi_2$ to the energy transfer rate of mode $\phi_1$ is equivalent to the contribution of the mode $\phi_2$ to scattering rate of the mode $\phi_1$.

## 3. Results and Discussion

We investigate the intrinsic 3-phonon scattering rates ($\Gamma_{\phi_1}$) and the mode-to-mode scattering rates ($\Gamma_{\phi_1,\phi_2}$) for all 504 branches in RDX for a uniform $6 \times 6 \times 6$ k-points to identify the modes that are responsible for the majority of energy transfer to the NN bond distortion modes i.e., bands IV, VII and IX. Due to crystal symmetry of RDX, the 216 k-points reduce to a total of 64 unique k-points, which result in a total of $504 \times 64 = 32256$ phonon modes. The calculated 3-phonon scattering rates and phonon lifetimes are shown in Figure 1 and the magnitude of modewise Gruneisen parameter and 3-phonon phase space volume for all modes in RDX are shown in Figure 2. The low frequency molecular translation modes up to 90 cm$^{-1}$ have an average scattering rate of 233.21 ps$^{-1}$ and an average relaxation time of 6.3 fs. Such high scattering rates are due to the strong anharmonic coupling indicated by the large Gruneisen parameter values, and due to a large number of 3-phonon scattering events indicated by the large phase space volumes. The number of modes, the average scattering rate and relaxation time for each band, and the highest and lowest relaxation time within each band are shown in Table 2.

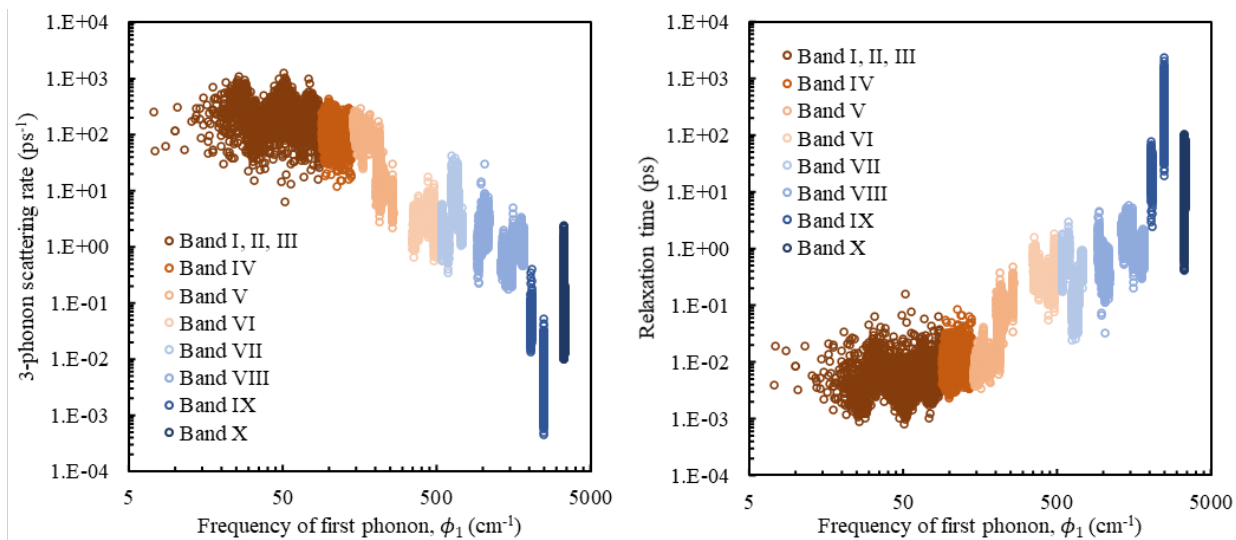

**Figure 1.** FGR based (left) 3-phonon scattering rate ($\Gamma_{\phi_1}$, ps$^{-1}$) and (right) phonon lifetime ($\tau_{\phi_1}$, ps) for all modes in RDX. Band I, II, III modes have the highest scattering rates and the shortest phonon lifetimes.

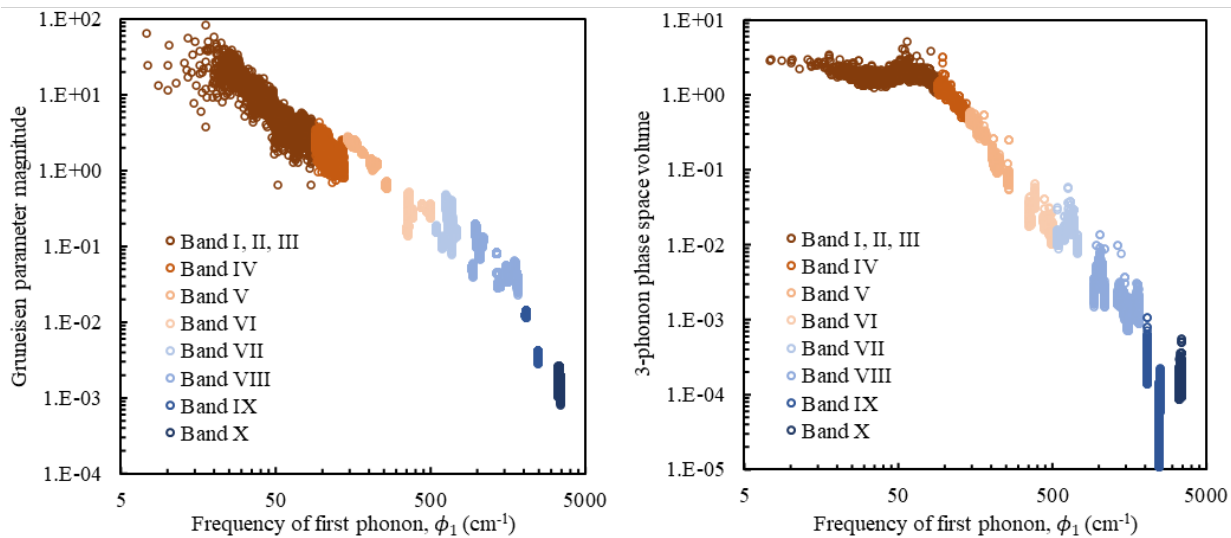

**Figure 2.** (left) Modewise Gruneisen parameter and (right) 3-phonon phase space volume for all modes in RDX. High scattering rates of bands I, II and III in Figure 1 are due to the strong anharmonic coupling indicated by the large Gruneisen parameter values, and due to a large number of 3-phonon scattering events indicated by the large phase space volumes.

| Bands | No. of modes | Average scattering rate (ps$^{-1}$) | Average relaxation time (ps) | Highest relaxation time (ps) | Lowest relaxation time (ps) |
| --- | --- | --- | --- | --- | --- |
| I | 325 | 310.01 | 0.005 | 0.045 | 0.001 |
| II | 727 | 221.22 | 0.007 | 0.068 | 0.001 |
| III | 1667 | 223.60 | 0.006 | 0.159 | 0.001 |
| IV | 1889 | 127.13 | 0.010 | 0.086 | 0.002 |
| V | 2560 | 48.73 | 0.065 | 0.466 | 0.003 |
| VI | 3072 | 2.91 | 0.423 | 1.797 | 0.056 |
| VII | 5120 | 6.80 | 0.273 | 2.902 | 0.024 |
| VIII | 10752 | 1.36 | 1.148 | 5.638 | 0.033 |
| IX | 3072 | 0.03 | 172.550 | 2272.898 | 2.457 |
| X | 3072 | 0.17 | 25.382 | 101.416 | 0.408 |

**Table 2.** Number of modes, average scattering rate and relaxation time for each band, and highest and lowest relaxation time within each band.

These results are consistent with the vibrational energy transfer dynamics reported by Ramasesha et al. who used ultrafast infrared spectroscopy for thin film RDX [48]. Following the excitation of a narrow band at 1533 cm$^{-1}$, instantaneous vibrational energy transfer (within 200 fs) to all modes was observed, indicating the strong anharmonic coupling in RDX. This instantaneous energy transfer could be due to the high scattering rate of the low frequency modes (0 to 90 cm$^{-1}$) which transfer the majority of the vibrational energy to all the modes within 159 fs as shown in Table 2 and Table 3. The authors also observed energy transfer dynamics up to 10 ps for modes in the 800 to 900 cm$^{-1}$ and 1200 to 1600 cm$^{-1}$ regions which are consistent with the relaxation time of band VIII modes of our results. The evolution of vibrational energy redistribution for NO$_2$ stretching, NN stretching and CN stretching modes were observed for over 100 ps. This could be due to the relatively weaker anharmonic coupling and lower number of 3-phonon scattering events for these modes (band IX of our results) as shown in Figure 2 which result in relaxation times as high as 2273 ps. Although, the average Gruneisen parameter value of the modes in band IX is higher than band X, however, band IX modes have lower scattering rates due to much lower number of allowable 3-phonon scattering events (smaller phase space volume $P_{3,\phi_1}$). The scattering rate of individual modes ($\Gamma_{\phi_1}$) indicates the time scale for relaxation of the phonon population of that mode. However, mode-to-mode scattering rates are needed to identify the modes that are responsible for the majority of energy transfer into or out of any particular mode of interest. The mode-to-mode scattering rates ($\Gamma_{\phi_1,\phi_2}$) are shown in Figure 3 and the band-to-band scattering rate contributions ($\Gamma_{band1,band2}$) are shown in Table 3. These results indicate that the low frequency modes in bands I, II and III, which constitute ~8% of the total number of modes in RDX, are responsible for the majority of energy transfer to all the modes, accounting for 63.6% of the mode-to-mode scattering rates ($\Gamma_{\phi_1,\phi_2}$) averaged over all modes $\phi_1$, calculated as $\frac{1}{total\ no.of\ modes}\sum_{\phi_1} 100 \frac{\sum_{\phi_2 \in band\ I,II,III} \Gamma_{\phi_1,\phi_2}}{\Gamma_{\phi_1}}$. In contrast, the mid frequency bands IV through VIII, which constitute ~73% of the total number of modes, account for 32.8% of the mode-to-mode scattering and the high frequency bands IX and X, which constitute ~19% of the total number of bands, account for 3.6% of the mode-to-mode scattering averaged over all modes $\phi_1$. The trivial contribution of the high frequency modes to vibrational energy transfer is due to their weak anharmonic coupling and lack of participation in 3-phonon scattering as shown in Figure 2. The contribution of individual modes ($\phi_2$) to the scattering of NN bands ($\Gamma_{NN\ bands,\phi_2}$, bands IV, VII and IX correspond to the largest distortion of NN bonds and are referred to as NN bands) is shown in Figure 4. Bands I, II and III combined are responsible for ~89%, ~66% and ~52% of the scattering involving the modes in bands IV, VII and IX respectively. In contrast, the mid frequency bands IV through VIII combined are responsible for ~11%, ~34% and ~34% of the

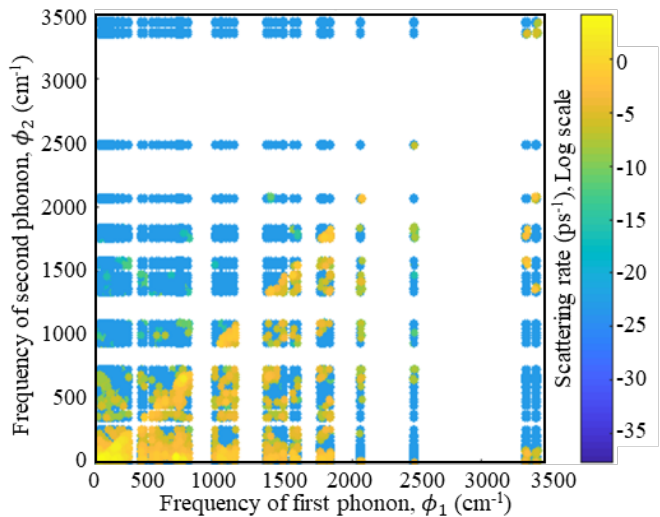

**Figure 3.** 3-phonon mode-to-mode scattering rates ($\Gamma_{\phi_1,\phi_2}$, ps$^{-1}$) for all modes in RDX. Low frequency modes up to 90 cm$^{-1}$ have the highest mode-to-mode scattering rates averaged over all modes $\phi_1$.

|  |  | Band 2 | | | | | | | | | |
|---|---|---|---|---|---|---|---|---|---|---|---|
|  |  | I | II | III | IV | V | VI | VII | VIII | IX | X |
| Band 1 | I | 69.25 | 19.25 | 8.46 | 1.91 | 0.40 | 0.09 | 0.15 | 0.31 | 0.09 | 0.09 |
|  | II | 70.00 | 20.05 | 8.05 | 1.69 | 0.20 | 0.00 | 0.01 | 0.00 | 0.00 | 0.00 |
|  | III | 48.02 | 37.95 | 13.10 | 0.88 | 0.05 | 0.01 | 0.00 | 0.00 | 0.00 | 0.00 |
|  | IV | 29.93 | 25.71 | 33.28 | 11.00 | 0.06 | 0.01 | 0.00 | 0.00 | 0.00 | 0.00 |
|  | V | 20.99 | 19.15 | 11.34 | 20.23 | 28.26 | 0.02 | 0.00 | 0.00 | 0.00 | 0.00 |
|  | VI | 24.44 | 13.82 | 18.85 | 13.01 | 9.33 | 20.29 | 0.25 | 0.01 | 0.00 | 0.00 |
|  | VII | 30.87 | 22.14 | 12.51 | 3.74 | 0.64 | 4.92 | 25.15 | 0.02 | 0.00 | 0.00 |
|  | VIII | 20.64 | 27.88 | 15.12 | 3.12 | 1.33 | 1.86 | 5.34 | 24.72 | 0.00 | 0.00 |
|  | IX | 52.05 | 0.00 | 0.01 | 0.01 | 0.16 | 0.26 | 15.38 | 18.48 | 13.63 | 0.02 |
|  | X | 8.30 | 0.00 | 33.45 | 1.99 | 0.00 | 0.00 | 0.00 | 31.67 | 6.04 | 18.55 |

**Table 3.** 3-phonon band-to-band scattering rate contributions ($\Gamma_{band1,band2}$, %). Rows represent $band1$ and columns represent $band2$. Each column represents the percent contribution of that band to the scattering rate of the phonons in the corresponding rows.

scattering of bands IV, VII and IX respectively. The highest frequency bands IX and X combined contribute negligibly to the scattering of bands IV and VII, but are responsible for

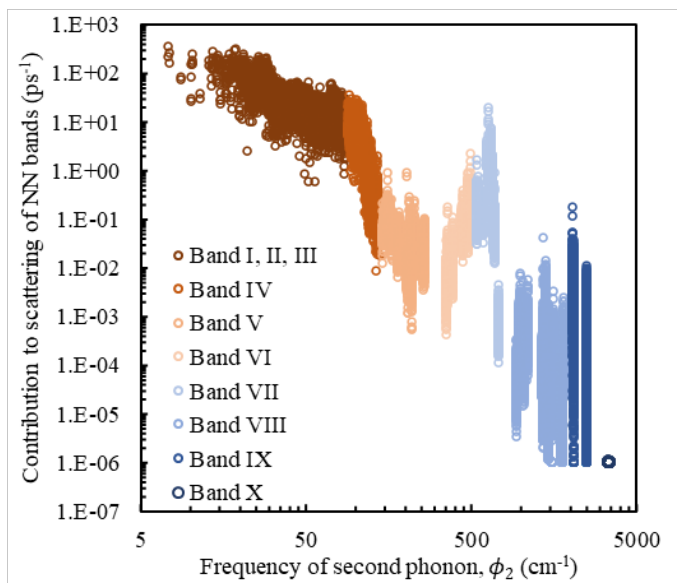

**Figure 4.** Contribution of individual modes to scattering of NN bands, $\Gamma_{NN\ bands,\phi_2}$ (ps$^{-1}$). Low frequency modes up to 90 cm$^{-1}$ in RDX are directly responsible for the majority of vibrational energy up-pumping to the largest NN bond distortion modes via 3-phonon scattering and mid-frequency doorway modes play a secondary role in energy up-pumping.

~14% of the scattering of band IX. Due to the proportionality between mode-to-mode scattering rates and mode-to-mode energy transfer rates derived in Eq. (13), these results indicate that the low frequency modes up to 90 cm$^{-1}$ in RDX are directly responsible for the majority of vibrational energy transfer to the largest NN bond distortion modes via 3-phonon scattering and the mid frequency modes play a secondary role in vibrational energy up-pumping. This is in contrast to the belief that the majority of the energy up-pumping to NN bond distortion modes takes place through an indirect route involving the mid frequency doorway modes where the doorway modes are responsible for the majority of scattering of the NN bands.

Based on these observations, a hybrid mechanism of energy up-pumping via both direct and indirect pathways is likely where the direct energy transfer dominates but a nontrivial amount of energy is transferred through an indirect route. A large number of 3-phonon scattering channels is available to the low frequency molecular translation modes up to 90 cm$^{-1}$ that are highly anharmonic and therefore scatter quickly and redistribute the majority of energy to all the modes including the largest NN bond distortion modes within 159 fs. Similar fast energy transfer dynamics were observed by Filipiak et al. [29] who reported vibrational energy transfer to all the modes in PETN within 200 fs following excitation at 1660 cm$^{-1}$. This could have resulted from the quick vibrational energy transfer to the highly anharmonic low frequency modes which then transfer the vibrational energy to all the modes via a large number of 3-phonon scattering channels. In addition, the transient spectra of the modes in PETN presented in [29] show strong absorption within 1 ps timescale. Such fast energy transfer dynamics which are commensurate with the sub-picosecond relaxation times of the low frequency molecular translation modes indicate the primary role played by these modes in direct vibrational energy transfer to all the modes including the NN bands. Although the quantitative contribution of the low frequency modes in energy transfer is not well-studied, our results indicate that among the low frequency modes, band I modes which are the lowest frequency modes transfer more energy to all the modes than band II or III. Averaged over all first phonon modes ($\phi_1$), the contribution of band I to mode-to-mode energy transfer is 1.88 times the contribution of band II and 3.09 times the contribution of band III. The direct energy transfer from these low frequency modes lasts for less than 0.16 ps which is the highest relaxation time of phonons in the frequency range 0 to 90 cm$^{-1}$. Next, the mid frequency modes from 90 to 1839 cm$^{-1}$, which include the largest ring bending, nitro and nitramine distortion modes, further up-pump the energy to the NN bands. Among the mid frequency modes, band VII and VIII modes transfer more energy to the NN bands than band IV, V or VI. This indirect energy transfer from the mid frequency modes lasts up to 5.6 ps. By this point, the majority of energy up-pumping to the NN bands is complete leading to an increase in the phonon population and large amplitude of internal molecular vibrations which may result in bond dissociation. Finally, the high frequency modes from 2055 to 3444 cm$^{-1}$ scatter and redistribute a small fraction of the vibrational energy to all other modes. The energy transfer from these high frequency modes is slow and lasts for over 2000 ps.

The vibrational energy transfer pathways via 3-phonon scattering identified in this work suggest that a unit of energy imparted to the lowest frequency modes under equilibrium would result in an instantaneous transfer of energy directly to all other vibrational modes including the majority of energy transfer to the critical NN bond distortion modes. A substantial amount of energy to the NN bonds is also transferred indirectly via the mid frequency doorway modes. Although the equilibrium conditions do not accurately imitate the material behavior under shock, qualitative conclusions be drawn regarding the strength of phonon-phonon scattering and the likely pathways of energy transfer which can be exploited to advance the design of new energetics and control the behavior of energetics during manufacturing, transport and use. The accuracy of the model can be improved at the cost of computational resources by including higher order anharmonic terms in the Hamiltonian (4-phonon scattering or higher) which become increasingly important at higher temperatures. Further improvement can be achieved by sampling the entire Brillouin zone using a greater number of kpoints, and by lifting the single mode relaxation time approximation to accurately model the change in phonon population and the anharmonic coupling between the modes. In addition, use of more accurate quantum mechanical models like Density Functional Theory to model the interaction between the atoms and molecules can provide greater insights into the physical and chemical processes that may lead to critical bond dissociation in energetics.

## 4. Conclusion

The modes corresponding to the largest distortion of NN bonds in RDX are identified (band IV, VII and IX) using a harmonic oscillator model. Next, an FGR based 3-phonon scattering model is used to calculate the mode-to-mode scattering rates. The low frequency molecular translation modes (between 0 and 90 cm$^{-1}$) dominate the energy transfer, accounting for ~89%, ~66% and ~52% of the scattering of band IV, VII and IX respectively. This indicates a dominant direct pathway for up-pumping of the vibrational energy from the low frequency molecular translation modes to the NN bond distortion modes. Besides the direct energy transfer, the modes from 90 to 1839 cm$^{-1}$ also contribute significantly to the transfer of energy to the NN bands. The highest frequency modes have the lowest contribution to energy transfer due to their trivial participation in phonon-phonon scattering. Based on these observations, a hybrid mechanism of energy up-pumping via both direct and indirect pathways is likely where the direct energy transfer dominates but a nontrivial amount of energy is transferred through an indirect route. The low frequency modes up to 90 cm$^{-1}$ scatter quickly and redistribute the energy to all the modes within 0.16 ps. Next, the mid frequency modes from 90 to 1839 cm$^{-1}$ further up-pump the energy to the NN bands within 5.6 ps. Finally, the high frequency modes from 2055 to 3444 cm$^{-1}$ scatter and redistribute a small fraction of the vibrational energy to all other modes which lasts over 2000 ps. Based on these observations, we believe that the very low frequency modes which scatter quickly and contribute the most to the transfer of vibrational energy are the most critical for phenomena leading to initiation in energetics.

## Supporting Information

- Theoretical and computational details for estimating the bond distortion due to phonon modes, and calculating the scattering rates.

## Author Information

**Corresponding Author**

* Email: pchung15@umd.edu

**Author Contributions**

The manuscript was written through contributions of all authors. All authors have given approval to the final version of the manuscript.

## Acknowledgment

G. K. gratefully acknowledges the graduate fellowship from the Center for Engineering Concepts Development and the Kulkarni Foundation Summer Research Fellowship. This work was also supported, in part, by the Department of Mechanical Engineering at the University of Maryland College Park.

## References

[1] F. Bowden and A. Yoffe, "Hot Spots and the Initiation of Explosion," *Symp. Combust. Flame Explos. Phenom.,* vol. 3, no. 1, pp. 551-560, 1949.

[2] F. Bowden and A. Yoffe, Fast Reactions in Solids, London: Butterworths Scientific Publications, 1958.

[3] F. Bowden and A. Yoffe, Initiation and Growth of Explosion in Liquids and Solids, Cambridge: Cambridge University Press, 1952.

[4] D. I. Bolef and M. Menes, "Nuclear Magnetic Resonance Acoustic Absorption in KI and KBr," *Phys. Rev.,* vol. 114, no. 6, pp. 1441-1451, 1959.

[5] A. W. Campbell, W. C. Davis, J. B. Ramsay and J. R. Travis, "Shock Initiation of Solid

Explosives," *Phys. Fluids,* vol. 4, no. 4, pp. 511-521, 1961.

[6] C. L. Mader, "Shock and Hot Spot Initiation of Homogeneous Explosives," *Phys. Fluids,* vol. 6, no. 3, pp. 375-381, 1963.

[7] F. Zerilli and E. Totton, "Shock-Induced Molecular Excitation in Solids," *Phys. Rev. B,* vol. 29, no. 10, p. 5891–5902, 1984.

[8] C. Coffey and E. Totton, "A Microscopic Theory of Compressive Wave-Induced Reactions in Solid Explosives," *J. Chem. Phys.,* vol. 76, no. 2, p. 949–954, 1982.

[9] F. E. Walker, "Physical kinetics," *J. Appl. Phys.,* vol. 63, no. 11, pp. 5548-5554, 1988.

[10] D. D. Dlott, "Optical Phonon Dynamics in Molecular Crystals," *Annu. Rev. Phys. Chem.,* vol. 37, no. 1, p. 157, 1986.

[11] W. L. Wilson, G. Wäckerle and M. D. Fayer, "Impurity perturbed domains: Resonant enhancement of bulk mode CARS by coupling to the electronic states of dilute impurities," *J. Chem. Phys.,* vol. 87, no. 5, pp. 2498-2504, 1987.

[12] J. R. Hill, E. L. Chronister, T. C. Chang, H. Kim, J. C. Postlewaite and D. D. Dlott, "Vibrational relaxation and vibrational cooling in low temperature molecular crystals," *J. Chem. Phys.,* vol. 88, no. 2, pp. 949-967, 1988.

[13] D. D. Dlott and M. D. Fayer, "Shocked moelcular solids: Vibrational up pumping, defect hot spot formation, and the onset of chemistry," *The Journal of Chemical Physics,* vol. 92, no. 6, pp. 3798-3812, 1990.

[14] D. D. Dlott, "New Developments in the Physical Chemistry of Shock Compression," *Annual Review of Physical Chemistry,* vol. 62, no. 1, pp. 575-597, 2011.

[15] A. Tokmakoff, M. D. Fayer and D. D. Dlott, "Chemical Reaction Initiation and Hot-Spot Formation in Shocked Energetic Molecular Materials," *Journal of Physical Chemistry,* vol. 97, no. 9, pp. 1901-1913, 1993.

[16] C. Aubuchon, K. Rector, W. Holmes and M. Fayer, "Nitro group asymmetric stretching mode lifetimes of molecules used in energetic materials," *Chemical Physics Letters,* vol. 299, no. 1, pp. 84-90, 1999.

[17] B. Kraczek and P. W. Chung, "Investigation of direct and indirect phonon-mediated bond excitation in α-RDX," *The Journal of Chemical Physics,* vol. 138, no. 7, p. 4505, 2013.

[18] L. E. Fried and A. J. Ruggiero, "Energy Transfer Rates in Primary, Secondary, and Insensitive Explosives," *J. Phys. Chem.,* vol. 98, p. 9786, 1994.

[19] S. Ye, K. Tonokura and M. Koshi, "Vibron dynamics in RDX, b-HMX and Tetryl crystals,"

*Chemical Physics,* vol. 293, no. 1, pp. 1-8, 2003.

[20] S. Ye, K. Tonokura and M. Koshi, "Energy Transfer Rates and Impact Sensitivities of Crystalline Explosives," *Combust Flame,* vol. 132, p. 240–246, 2003.

[21] S. Ye and M. Koshi, "Theoretical Studies of Energy Transfer Rates of Secondary Explosives," *The Journal of Physical Chemistry B,* vol. 110, no. 37, pp. 18515-18520, 2006.

[22] J. Bernstein, "Ab initio study of energy transfer rates and impact sensitivities of crystalline explosives," *J. Chem. Phys.,* vol. 148, no. 8, p. 084502, 2018.

[23] K. Joshi, M. Losada and S. Chaudhuri, "Intermolecular Energy Transfer Dynamics at a Hot-Spot Interface in RDX Crystals," *J. Phys. Chem. A,* vol. 120, no. 4, pp. 477-489, 2016.

[24] K. McNesby and C. Coffey, "Spectroscopic Determination of Impact Sensitivities of Explosives," *J. Phys. Chem. B,* vol. 101, no. 16, pp. 3097-3104, 1997.

[25] J. Ostrander, R. Knepper, A. Tappan, J. Kay, M. Zanni and D. Farrow, "Energy Transfer Between Coherently Delocalized States in Thin Films of the Explosive Pentaerythritol Tetranitrate (PETN) Revealed by Two-Dimensional Infrared Spectroscopy," *J. Phys. Chem. B,* vol. 121, no. 6, pp. 1352-1361, 2017.

[26] G. Yu, Y. Zeng, W. Guo, H. Wu, G. Zhu, Z. Zheng, X. Zheng, Y. Song and Y. Yang, "Visualizing Intramolecular Vibrational Redistribution in Cyclotrimethylene Trinitramine (RDX) Crystals by Multiplex Coherent Anti-Stokes Raman Scattering," *J. Phys. Chem. A,* vol. 121, no. 3, pp. 2565-2571, 2017.

[27] W. Wang, N. Sui, L. Zhang, Y. Wang, L. Wang, Q. Wang, J. Wang, Z. Kang, Y. Yang, Q. Zhou and H. Zhang, "Scanning the energy dissipation process of energetic materials based on excited state relaxation and vibration-vibration coupling," *Chinese Physics B,* vol. 27, no. 10, 2018.

[28] A. Michalchuk, S. Rudic, C. Pulham and C. Morrison, "Vibrationally induced metallisation of the energetic azide α-NaN3," *Phys. Chem. Chem. Phys.,* vol. 20, no. 46, pp. 29061-29069, 2018.

[29] N. Cole-Filipiak, R. Knepper, M. Wood and K. Ramasesha, "Sub-picosecond to Sub-nanosecond Vibrational Energy Transfer Dynamics in Pentaerythritol Tetranitrate," *Phys. Chem. Lett.,* vol. 11, no. 16, pp. 6664-6669, 2020.

[30] N. Cole-Filipiak, M. Marquez, R. Knepper, R. Harmon, D. Wiese-Smith, P. Schrader, M. Wood and K. Ramasesha, "Ultrafast spectroscopic studies of vibrational energy transfer in energetic materials," *AIP Conference Proceedings,* vol. 2272, no. 1, p. 060006, 2020.

[31] M. P. Kroonblawd, T. D. Sewell and J. B. Maillet, "Characteristics of energy exchange between inter- and intramolecular degrees of freedom in crystalline 1,3,5-triamino-2,4,6-trinitrobenzene (TATB) with implications for coarse-grained simulations of shock waves in

polyatomic molecular crystals," *J. Chem. Phys.,* vol. 144, no. 6, p. 064501, 2016.

[32] M. P. Kroonblawd and L. E. Fried, "High Explosive Ignition through Chemically Activated Nanoscale Shear Bands," *PRL,* vol. 124, no. 20, p. 206002, 2020.

[33] E. Fermi, Nuclear Physics, Chicago: University of Chicago Press, 1950.

[34] J. Hooper, "Vibrational energy transfer in shocked molecular crystals," *The Journal of Chemical Physics,* vol. 132, p. 014507, 2010.

[35] Y. Long and J. Chen, "Theoretical study of the phonon–phonon scattering mechanism and the thermal conductive coefficients for energetic material," *Philosophical Magazine,* vol. 97, no. 28, pp. 2575-2595, 2017.

[36] A. Michalchuk, M. Trestman, S. Rudic, P. Portius, P. Fincham, C. Pulham and C. Morrison, "Predicting the reactivity of energetic materials: an ab initio multi-phonon approach," *J. Mater. Chem. A,,* vol. 7, p. 19539, 2019.

[37] C. L. Schosser and D. D. Dlott, "A picosecond CARS study of vibron dynamics in molecular crystals: Temperature dependence of homogeneous and inhomogeneous linewidths," *J. Chem. Phys.,* vol. 80, no. 4, pp. 1394-1406, 1984.

[38] G. Kumar, F. G. VanGessel, D. C. Elton and P. W. Chung, "Phonon Lifetimes and Thermal Conductivity of the Molecular Crystal α-RDX," *MRS Advances,* pp. 1-9, 2019.

[39] G. Kumar, F. G. VanGessel and P. W. Chung, "Bond Strain and Rotation Behaviors of Anharmonic Thermal Carriers in a-RDX," *Propellants, Explosives, Pyrotechnics,* vol. 45, no. 2, 2019.

[40] F. G. VanGessel, G. Kumar, D. C. Elton and P. W. Chung, "A Phonon Boltzmann Study of Microscale Thermal Transport in α-RDX Cook-Off," *arXiv:1808.08295,* pp. 1-11, 2018.

[41] L. B. Munday, P. W. Chung, B. M. Rice and S. D. Solares, "Simulations of high-pressure phases in RDX," *The Journal of Physical Chemistry B,* vol. 115, no. 15, pp. 4378-4386, 2011.

[42] G. D. Smith and R. K. Bharadwaj, "Quantum Chemistry Based Force Field for Simulations of HMX," *The Journal of Physical Chemistry B,* vol. 103, no. 18, pp. 3570-3575, 1999.

[43] J. D. Gale and A. L. Rohl, "The general utility lattice program (GULP)," *Molecular Simulation,* vol. 29, no. 5, pp. 291-341, 2003.

[44] J. D. Gale, "Analytical Free Energy Minimization of Silica Polymorphs," *J. Phys. Chem. B,* vol. 102, no. 28, pp. 5423-5431, 1998.

[45] S. Plimpton, "Fast Parallel Algorithms for Short-Range Molecular Dynamics," *J Comp*

*Phys,* vol. 117, 1995.

[46] W. Li, J. Carette, N. A. Katcho and N. Mingo, "ShengBTE: A solver of the Boltzmann transport equation for phonons," *Computer Physics Communications,* vol. 185, no. 6, pp. 1747-1758, 2014.

[47] R. I. Castillo, L. P. Londono and S. P. H. Rivera, "Vibrational spectra and structure of RDX and its 13C- and 15N-labeled derivatives: A theoretical and experimental study," *Spectrochimica Acta Part A: Molecular and Biomolecular Spectroscopy,* vol. 76, no. 2, pp. 137-141, 2010.

[48] K. Ramasesha, M. Wood, N. C. Filipiak and R. Knepper, "Experimental and Theoretical Studies of Ultrafast Vibrational Energy Transfer Dynamics in Energetic Materials," Sandia National Lab, Livermore, 2020.

[49] A. Werbin, "The Infrared Spectra of HMX and RDX," California Univ., Livermore, 1957.

[50] A. A. Maradudin and A. E. Fein, "Scattering of Neutrons by an Anharnmnic Crystal," *Physical Review,* vol. 128, no. 6, pp. 2589-2608, Dec 1962.

[51] A. A. Maradudin, A. E. Fein and G. H. Vineyard, "On the evaluation of phonon widths and shifts," *physica status solidi (b),* vol. 2, no. 11, pp. 1479-1492, 1962.

[52] R. Peierls, Quantum theory of solids, London: Oxford University Press, 1955.

**For Table of Contents Only**

There is a significant error in this paper. Please refer to the corrected version that is now available in the corresponding journal article https://pubs.acs.org/doi/10.1021/acs.jpca.1c03225

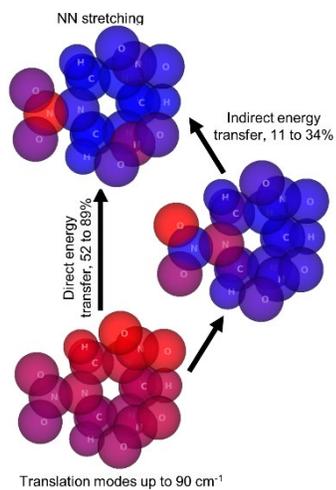